\documentclass[english,aps,pra,twocolumn,superscriptaddress]{revtex4-1}
\usepackage[utf8]{inputenc}	
\usepackage{geometry}		
\usepackage{graphicx}
\usepackage[above,below]{placeins}	
\usepackage{times}
\usepackage{dcolumn}
\usepackage{bm}
\usepackage{enumerate}
\usepackage{enumitem}
\usepackage{ctable}
\usepackage{booktabs}
\usepackage{natbib}
\usepackage{siunitx}
\usepackage{pifont}
\usepackage{csquotes}

\begin{document}

\title{What are we assessing? An analysis of the most common concept inventories in physics}

\author{James T. \surname{Laverty}}
\affiliation{Department of Physics, Kansas State University, Manhattan, KS 66506, USA}

\author{Marcos D. \surname{Caballero}}
\affiliation {Department of Physics and Astronomy \& CREATE for STEM Institute, Michigan State University, East Lansing, Michigan 48824}
\affiliation {Department of Physics and Center for Computing in Science Education, University of Oslo, N-0316 Oslo, Norway}


\begin{abstract}
  Assessing student learning is a cornerstone of educational practice. Standardized assessments have played a significant role in the development of instruction, curricula, and educational spaces in college physics. However, the use of these assessments to evaluate student learning is only productive if they continue to align with our learning goals. Recently, there have been calls to elevate the process of science (``scientific practices'') to the same level of importance and emphasis as the concepts of physics (``core ideas'' and ``crosscutting concepts'').  We use the recently developed 3D-LAP to investigate how well the most commonly used standardized assessments in introductory physics (i.e. concept inventories) align with this modern understanding of physics education's learning goals. We find that many of the questions on concept inventories do elicit evidence of student understanding of core ideas, but do not have the potential to elicit evidence of scientific practices or crosscutting concepts. Furthermore, we find that the individual scientific practices and crosscutting concepts that are assessed using these tools are limited to a select few. We discuss the implications that these findings have on designing and testing curricula and instruction both in the past and for the future.
\end{abstract}

\maketitle

\section{Introduction}\label{introduction}

Assessment helps us to understand what students know and are able to do after instruction; it aids us in understanding which aspects of a curriculum are working well for students and which are not; and it provides us with evidence of how well students are meeting our intended learning outcomes \cite{national_research_council_knowing_2001}. Taken as evidence of learning in physics, different forms of standardized assessment have helped shape many of the major changes that have occurred in physics education over the last 40 years \cite{national_research_council_adapting_2013,mcdermott_resource_1999,meltzer_resource_2012}. Standardized assessment practices in undergraduate physics education emphasize the use of conceptual pre- and post-tests (``concept inventories'') -- the outcomes of which have been used to inform changes to curriculum design and instructional practices \cite{madsen_resource_2016}. A wide variety of studies have been conducted using concept inventories \cite{pollock_sustaining_2008, beichner_case_1999,  redish_effectiveness_1997, sorensen_new_2006, lorenzo_reducing_2006, hoellwarth_direct_2005, lasry_peer_2008, prather_national_2009} and student learning outcomes on such assessments are well documented \cite{etkina_ISLE_chapter, crouch_peer_2001, crouch_Peer_Instruction_chapter, beichner_SCALEUP_chapter, chabay_sherwood_MandI_chapter, hake_interactive-engagement_1998, freeman_active_2014}.

At the same time we ask these questions, physics education researchers, curriculum developers, and instructors have used the outcomes of concept inventories to inform their work.
But what are these inventories assessing? What learning goals were used to inform their design? And how well might these concept inventories represent an assessment of the learning outcomes in typical physics courses?

Physics education research has begun to address a wider variety of learning outcomes over the years \cite{national_research_council_adapting_2013, kozminski_aapt_2014,behringer2017aapt}. Courses that were once focused heavily on conceptual understanding, now include engagement in scientific practice, the development of more sophisticated epistemologies, and achieving positive attitudinal shifts towards physics. Curriculum design literature argues that aligning assessments and instruction with these goals is critical to helping students achieve these goals (e.g., ``backwards design''). In particular, the assessments we use are meant to develop an evidentiary argument for student learning  \cite{wiggins_understanding_2005, biggs_enhancing_1996, national_research_council_knowing_2001}. Arguably, the common concept inventories in physics are insufficient to address these broader learning outcomes. We are saddled with tools that provide some information, but this information is becoming increasingly incomplete for researchers, curriculum developers, and instructors. It is reasonable to ask: what do our current assessments tell us about student learning? That is, what are we assessing?

In this paper, we address these questions using the framework of three-dimensional learning (3DL)  \cite{national_research_council_framework_2012}. While this lens backgrounds a number of important issues (e.g., epistemological development, and shifts in identity), it foregrounds engaging students in the process of science (scientific practices) and helping students develop how they organize their knowledge (core ideas and crosscutting concepts). Our analysis makes use of the recently developed Three-Dimensional Learning Assessment Protocol (3D-LAP)  \cite{laverty_characterizing_2016} --  a tool that evidences how well assessments provide opportunities to engage students in 3DL. Using the 3D-LAP, we coded the questions appearing on the four most common concept inventories (FCI, FMCE, BEMA, CSEM) to determine the degree to which they can provide evidence of 3DL. This paper provides a brief discussion of standardized assessment in physics (Sec.~\ref{standardized}), offers an overview of 3DL (Sec.~\ref{learninggoals}) with more details in Ref. \cite{national_research_council_framework_2012}, reviews the 3D-LAP (Sec.~\ref{3DLAP}) and its use to analyze assessment tasks (Sec. ~\ref{methods}), but defers to Ref. \cite{laverty_characterizing_2016} for details, and analyzes the four most common standardized assessments in physics using the 3DLAP (Sec.~\ref{results}). We provide concluding remarks in Sec.~\ref{conclusion}.

\section{Standardized Assessment in Physics}\label{standardized}

Standardized assessment is widely used in physics education to measure learning outcomes in a variety of physics courses \cite{etkina_ISLE_chapter, crouch_peer_2001, beichner_case_1999, chabay_sherwood_MandI_chapter} including, most recently, upper-division courses \cite{caballero_assessing_2014, chasteen2012colorado, wuttiprom2009development, goldhaber2009transforming, mckagan2006exploring, cataloglu2001development}. It is typical to use these standardized assessments as ``summative assessments'' for a given course where they are used to gather evidence of what students have learned at the time that they take them, with little intention of using them to help those same students learn physics. That is, we typically assume (even without being explicit about it) that concept inventories attempt to elicit, identify, and track stable cognitive elements. Because of that stability, we neglect any learning that occurs during the assessment itself  \cite{sayre_plasticity:_2007, sayre_plasticity_2008}. Some learning may occur when students interact with the measurement tool, but those effects are assumed to be small compared to the learning that has occurred over the time period that people are trying to measure (i.e., one semester) \cite{vygotsky1980mind}.

Concept inventories have typically focused on measuring ``conceptual change'' or ``expert-like thinking.'' Their development has varied, but often follows a similar procedure \cite{adams_development_2011}.
This process usually starts by developing a large number of questions around the target concept -- using the current literature on common misconceptions or difficulties around that concept as a guide.  These initial questions are usually open-ended and are presented to the target audience (students) under test conditions, in think-aloud interviews, or both.  The developers then use the students' responses to eliminate or to modify questions that do not meet their standards (e.g. students did not interpret the question as intended or almost everyone got the question right). In addition, developers pay attention to common student responses to the questions. The questions that are deemed appropriate are then converted into multiple-choice questions where the distractor answer options match these most common incorrect responses. For open-ended assessments, it is common for the grading rubric to include the most common incorrect responses \cite{caballero_assessing_2014, chasteen2012colorado}. The test is re-administered to students and modified as necessary until the developers are satisfied with the results. These results might be achieving some sort of stability in student performance, some set of appropriate test statistics, or both. Here, we do not intend to suggest that the development of concept inventories is straight-forward or simple; it is not. There is certainly nuance in the design and development of specific inventories. However, the general process described above is quite similar to the development of the commonly-used concept inventories in introductory physics.

The Force Concept Inventory (FCI) is almost certainly the most well known and widely used standardized assessment in introductory physics courses  \cite{hestenes_force_1992}. Both it and the Force and Motion Conceptual Evaluation (FMCE) are designed to evaluate student learning of topics commonly found in the first semester of an introductory physics sequence  \cite{thornton_assessing_1998}. Similarly, both the Conceptual Survey of Electricity and Magnetism (CSEM) and the Brief Electricity and Magnetism Assessment (BEMA) were developed to evaluate student learning of topics commonly taught in the second semester of an introductory physics sequence  \cite{maloney_surveying_2001,ding_evaluating_2006}.

These (and other) concept inventories have provided straightforward, off-the-shelf ways to evaluate instructional practices and curricular materials \cite{madsen_research-based_2016}. Because of this, they have been used routinely to evaluate student learning in interactive environments \cite{hake_interactive-engagement_1998,freeman_active_2014, brewe_toward_2010}, to compare student learning in different environments \cite{kohlmyer_tale_2009,caballero_comparing_2012}, and to investigate different learning outcomes for different groups of students within classes \cite{madsen_gender_2013, brewe_toward_2010, lorenzo_reducing_2006}.
Using concept inventories in this way aligns with backward design; evidence should be collected to determine if instruction and curricula are helping all students achieve the learning goals we have for them.
However, standardized assessments that gather evidence of student learning are only useful if they align with our learning goals.  Recently, national reports have highlighted new ways to think about what we want our students to learn, both in K-12 and undergraduate science education.  In particular, these reports have emphasized the idea of blending the concepts, on which concept inventories have been focused, and practices of science together into our learning goals \cite{national_research_council_discipline-based_2012, kozminski_aapt_2014, college_board_ap_2015}.

\section{Evolving Learning Goals}\label{learninggoals}

Recent national calls have emphasized the need for students to engage with science and engineering practices at the same level of emphasis as they engage with science concepts \cite{national_research_council_framework_2012, national_research_council_discipline-based_2012}.
Changes to courses aligned with these calls broaden the scope of the learning goals in traditional introductory and advanced science courses and, as such, broaden the space for assessment. In physics, discussions of important practices have appeared in the revised advanced placement curriculum \cite{college_board_ap_2015} and in white papers describing the need for new laboratory and computational experiences for physics students \cite{kozminski_aapt_2014,behringer_aapt_2016}.

One national report aimed to synthesize the years of research on student learning in science courses into recommendations for curricula and instruction. {\it A Framework for K-12 Science Education: Practices, Crosscutting Concepts, and Core Ideas} (herein referred to as ``{\it Framework}''), gives a comprehensive view of merging the concepts and process of science  \cite{national_research_council_framework_2012}. The underlying idea of the {\it Framework} is that having students engage in science in the manner that scientists do while using scientific knowledge is a more productive way to build students' understanding of both the process and knowledge of science. By focusing on blending concepts and practices together, we aim to provide our students with a deeper, richer, more enduring learning experience that are likely to benefit both their epistemological and identity development (even though these ideas are backgrounded by 3DL). To be clear, this is not the idea that we must ``sacrifice'' the content to make room for the process of science; it is that the concepts and practices {\it are} the content. While the {\it Framework} was written for the K-12 education system, it has been argued that these ideas are relevant to higher education  \cite{cooper_challenge_2015,national_research_council_discipline-based_2012, mcdonald_next_2015}.

In this paper, we will use the ideas highlighted in the {\it Framework} as the basis for our analysis of the concept inventories to investigate how well our current assessments can provide evidence of learning of these broader goals. The {\it Framework} divides what we want students to learn into three ``dimensions'' of learning,  one that is practice-focused and two that are concept-focused. A brief description of each of the three dimensions is given here along with an example. We encourage the reader to look at the {\it Framework} if they are interested in deeper explanations of the dimensions  \cite{national_research_council_framework_2012}.

\begin{description}[noitemsep]
\item[Scientific Practices] These are the disaggregated components of the process of science. They involve putting scientific knowledge to use to model, predict, and explain phenomena (e.g., {\it Developing and Using Models}).
\item[Crosscutting Concepts] These bridge the boundaries between the disciplines of the physical, biological, and geological sciences. These ``ways of thinking'' are used by each discipline and can be leveraged to help students make connections across the sciences and between their classes (e.g., {\it Systems and System Models}).
\item[Disciplinary Core Ideas] These are the foundational concepts that are fundamental to the scientific discipline.  In order to qualify as a disciplinary core idea, the concept must 1) be essential to the study of the discipline, 2) be required to explain a wide range of phenomena, and 3) provide a way to generate new ideas and predictions (e.g., {\it Energy}).
\end{description}

The {\it Framework} emphasizes that it is vital that all three of these dimensions are blended into instruction, curriculum, and (most importantly for this article) assessments. Herein, we refer to the blending of these ideas as ``three-dimensional learning.''

In physics, we often use concept inventories to assess the outcomes in our courses (Sec.~\ref{standardized}), but how well do these inventories represent our shifting goals? In particular,

\begin{enumerate}
    \item How well do the four most commonly used concept inventories for introductory physics assess the goals of three-dimensional learning?
    \item For which, if any, of the Scientific Practices, Crosscutting Concepts, and Core Ideas, do these concept inventories provide some evidence of student learning?
\end{enumerate}

Note that the concept inventories that we are analyzing were developed well before the idea of three-dimensional learning. We understand that holding them to the standard that they should assess three-dimensional learning is not entirely fair.  However, our goal here is not to disparage these assessments. They provide important information regarding conceptual learning in many courses and have helped advance PER in substantial ways. Instead, we aim to survey the current state of standardized assessment in physics education and use this as a step towards discussing the next generation of standardized assessments.

\section{The Three-Dimensional Learning Assessment Protocol}\label{3DLAP}

In 2014, the NRC released a document highlighting the importance and challenges of developing assessments for the NGSS and (more broadly) three-dimensional learning \cite{national_research_council_developing_2014}. To help identify and develop assessments that are capable of eliciting evidence of students engaging with each of the three dimensions, we developed the Three-Dimensional Learning Assessment Protocol (3D-LAP) \cite{laverty_characterizing_2016}.

The 3D-LAP was designed with two central purposes:
\begin{enumerate}
\item To help researchers characterize how well assessments align with each of the three dimensions.
\item To help instructors develop or modify existing assessment tasks so that they have the potential to elicit evidence of students engaging with the three dimensions.
\end{enumerate}

The 3D-LAP uses individual questions or clusters of related questions (referred to herein as a ``task'') as the unit of analysis. By analyzing only the task itself, the 3D-LAP can be used to determine if the task has the potential to elicit evidence that a student will engage in a scientific practice, crosscutting concept, or core idea \cite{laverty_characterizing_2016}.

The 3D-LAP was developed as part of a larger project to transform the introductory physics, chemistry, and biology courses at Michigan State University. The development team (made up of the authors and 8 additional disciplinary experts, many of whom identify as discipline-based education researchers) initially developed a prototype set of criteria for each of the scientific practices, crosscutting concepts, and core ideas based on their descriptions in the {\it Framework}. Separately, we collected and discussed assessment tasks that exemplified each of the dimensions.  We then compared these exemplar tasks with the prototype criteria and used this comparison to revise and refine the criteria \cite{kolb2012grounded}. The final criteria took different forms for each dimension: scientific practices each have a list of 2-4 criteria, all of which must be met in order for a task to align with that scientific practice; crosscutting concepts each have a brief description of what is necessary to align with it; and each core idea comes with a list of ideas, at least one of which must be included in a task to qualify as aligning with a core idea.

Both the face and content validity of the 3D-LAP as applied to concept inventories is evidenced by the expertise of the development team. This team included disciplinary experts from physics, chemistry, and biology, some of whom identify as DBER and others that identify as more traditional experts. The development process reinforced the validity of the protocol by continually comparing the theory ({\it Framework}, research literature, etc.) and the on-the-ground reality (existing assessments). Some of these comparisons included assessment tasks from existing concept inventories in each of the disciplines.

In order to establish the reliability of the 3D-LAP when applied to these concept inventories, JTL coded all of the tasks, while MDC coded 25\% of the tasks chosen randomly. Cohen's Kappa is a commonly used measure of inter-rater reliability for two coders and it does well when looking for levels of agreement in many cases \cite{cohen_coefficient_1960}.  However, Cohen's Kappa does yield unexpected and uninformative values when the code appears in almost none (or almost all) of the cases, which is often the case when using the 3D-LAP \cite{cicchetti_high_1990}.  It is precisely because of these cases that Gwet's AC\textsubscript{1} was introduced \cite{gwet_computing_2008}.  Gwet's AC\textsubscript{1} is an alternative, more stable measure of agreement, even in cases where the codes appear very (in)frequently.

Our inter-rater reliability was established using Gwet's AC\textsubscript{1} statistic, obtaining a value of .93, .79, and .91 respectively for the scientific practices, crosscutting concepts, and core ideas \cite{gwet_computing_2008}. These values are typically considered good to very good agreement. For these purposes, we only check to see if both coders agreed that the task elicited a dimension or not, without regard to which component of the dimension was coded (i.e. if there is a scientific practice or not, not necessarily which scientific practice). This choice was made because we do not have the sufficient number of tasks needed to investigate the reliability of all 19 components of the 3D-LAP (7 scientific practices, 7 crosscutting concepts, and 5 core ideas). 

\section{Applying the 3D-LAP}\label{methods}

Here, we demonstrate how the 3D-LAP can be applied to assessment tasks in our data set; one that aligns with three-dimensional learning and one that does not. Because concept inventories require significant effort to develop and that effort can be compromised by making the inventories available to the public, we will not reprint any part of them here. Instead, we will describe two questions from the BEMA and refer the reader to the original exams for the exact questions \cite{ding_evaluating_2006}.

\subsection{Example 1: Alignment with one dimension}

Question 19 of the BEMA asks students about the difference in electric potential between any two points in a metal. The answer options all include a declaration of what that potential difference is, and a few words that are either about the value of the electric field (answer) or a common incorrect response.

Using the 3D-LAP, we characterize question 19 of the BEMA as providing no evidence that a student has engaged in a scientific practice or crosscutting concept, however, it does elicit the core idea of ``Interactions are Mediated by Fields''. The most closely associated scientific practice is Constructing Explanations and Engaging in Argument from Evidence. Column 2 of Table~\ref{table:SPcriteria} shows an analysis of the task to determine if it elicits this practice. As shown in Table~\ref{table:SPcriteria}, question 19 of the BEMA does ask the student to make a claim about the described situation, but does not present an event, observation, or phenomenon, or ask the student to select evidence or reasoning to support their claim. A student certainly might engage in the practice, but the question as written does not provide any evidence that they are being asked to do so. Similarly, this task does not elicit any of the crosscutting concepts as determined by the 3D-LAP.  The most closely associated crosscutting concept is Cause and Effect: Mechanism and Explanation.  The 3D-LAP criteria for this crosscutting concept is:
\begin{displayquote}
To code an assessment task with Cause and Effect: Mechanism and Explanation, the question provides at most two of the following: 1) a cause, 2) an effect, and 3) the mechanism that links the cause and effect, and the student is asked to provide the other(s).
\end{displayquote}
Question 19 of the BEMA does not ask the student to explain the mechanism that connects the cause to the effect. Unlike with the scientific practices and crosscutting concepts, the task does elicit evidence that a student has engaged with the core idea of ``Interactions are Mediated by Fields'', as the task asks the student specifically about the electric potential (and the correct answer includes the electric field).

\subsection{Example 2: Alignment with three dimensions}

Question 7 of the BEMA asks about the interactions between a charged object (wall) and a neutral object (rubber sheet). Each answer option includes a description of what will happen to the rubber sheet and a possible reason why. In contrast to Question 19, Question 7 of the BEMA does provide evidence that students can engage in a scientific practice, crosscutting concept, and core idea (at least as well as can be done in a multiple-choice question). Column 3 of Table~\ref{table:SPcriteria} shows the analysis of this task and gives a brief explanation of why it does align with the criteria for the scientific practice of Constructing Explanations and Engaging in Argument from Evidence. This task also elicits a crosscutting concept: Structure and Function. The 3D-LAP criteria for this crosscutting concept is:
\begin{displayquote}
To code an assessment task with Structure and Function, the question asks the student to predict or explain a function or property based on a structure, or to describe what structure could lead to a given function or property.
\end{displayquote}
Question 7 asks the student to use the atomic structure of the rubber sheet to predict the behavior of the sheet in response to the charged wall. Like Question 19, Question 7 also elicits evidence that a student has engaged with the core idea of ``Interactions are Mediated by Fields''.

\begin{table*}[htbp]
\caption{An analysis of Question 19 and Question 7 of the BEMA using the 3D-LAP criteria for the scientific practice, Constructing Explanations and Engaging in Argument from Evidence. An assessment task must meet all of the criteria in order for it to be considered to elicit that dimension.}
\begin{tabular}{ p{5.5cm} | p{5.5cm} | p{5.5cm}}

\textbf{3D-LAP criteria for aligning with Constructing Explanations and Engaging in Argument from Evidence} &\textbf{Characterization of BEMA question 19 with 3D-LAP criteria} & \textbf{Characterization of BEMA question 7 with 3D-LAP criteria} \\
1. Question gives an event, observation, or phenomenon. 	&\ding{55} 1. The question does not present a real-life situation (it takes place in an idealized model). &\ding{51} 1. This question is about a real-world scenario. \\
2. Question gives or asks student to select a claim based on the given event, observation, or phenomenon.	&\ding{51} 2. Question asks student to claim that the potential difference is zero or non-zero. &\ding{51} 2. Question asks student to claim whether or not the rubber sheet is affected by the wall. \\
3. Question asks student to select scientific principles or evidence in the form of data or observations to support the claim. 	&\ding{55} 3. Most answer options do not include scientific principles (charge, electric field).  &\ding{51} 3. Most answer options include scientific principles (charge, repulsion, polarization). \\
4. Question asks student to select the reasoning about why the scientific principles or evidence support the claim.	&\ding{55} 4. Answer options do not include the reasoning linking the principle and the claim. &\ding{51} 4. Most answer options include reasoning that connects the principles to the claim. \\

\end{tabular}
\label{table:SPcriteria}
\end{table*}

\section{Results of Coding Concept Inventories}\label{results}

Looking at the results of coding each question on a concept inventory in aggregate allows us to understand for which aspects of student learning the assessments are eliciting evidence. We have weighted the results of coding with the 3D-LAP using the percentage of points assigned to each question by the inventory authors to address our first research question: How well do the four most commonly used concept inventories for introductory physics assess the goals of three-dimensional learning? We first provide an overview and then discuss results for each concept inventory in turn.

\begin{figure}
\includegraphics[width=0.95\linewidth]{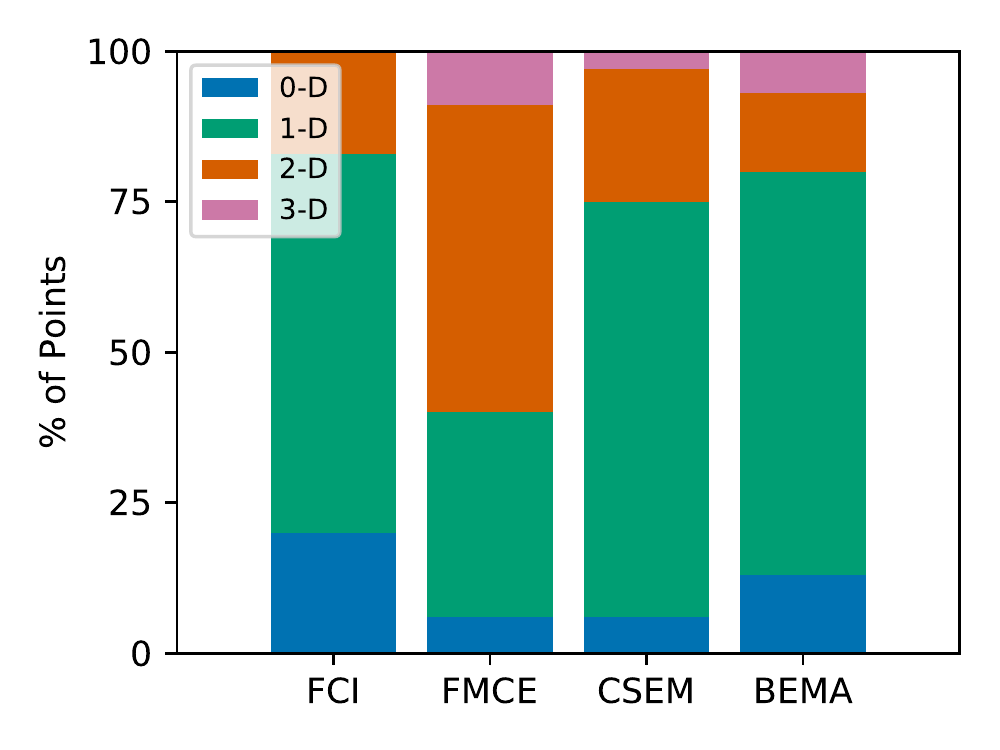}
\caption{Percentage of points for each concept inventory that have the potential to elicit evidence of 0, 1, 2, or all 3 dimensions.\label{figure:nDLpoints}}
\end{figure}

Fig.~\ref{figure:nDLpoints} shows that few of the tasks on the concept inventories address all three dimensions.  However, most of the tasks do assess at least one of the three dimensions, and few assess no dimensions. 

\begin{figure}
\includegraphics[width=0.95\linewidth]{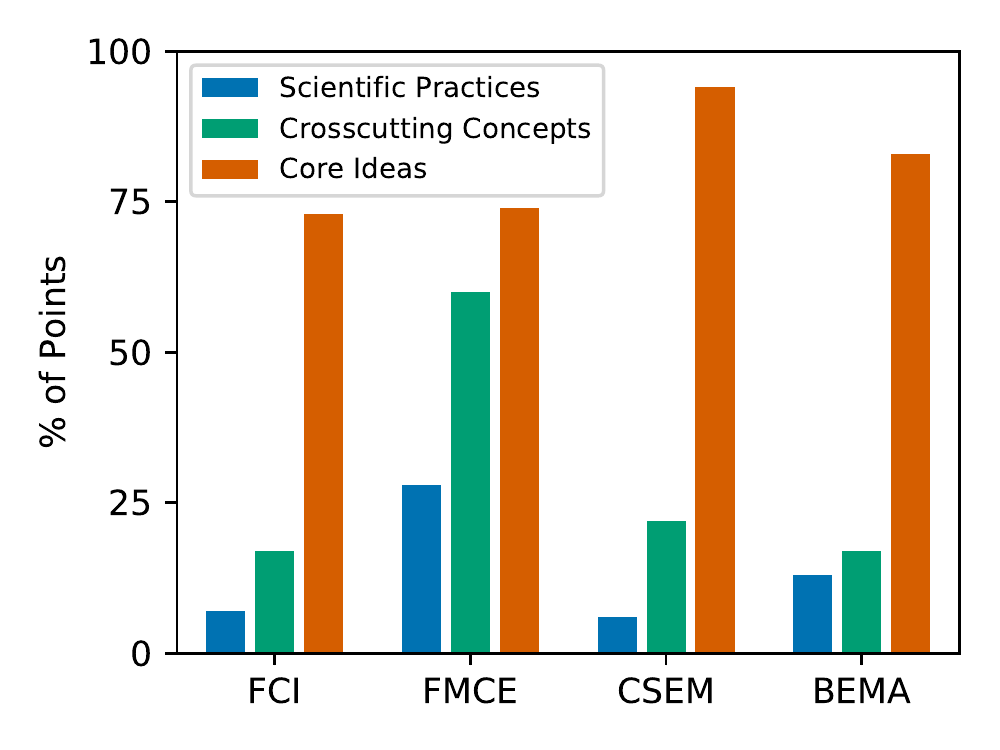}
\caption{Percentage of points for each concept inventory assigned to items that have the potential to elicit evidence of a scientific practice, crosscutting concept, or core idea.\label{figure:3DLpoints}}
\end{figure}

Fig.~\ref{figure:3DLpoints} provides a clearer picture of what the current concept inventories are assessing in terms of 3DL.  In each concept inventory, the majority of tasks have the potential to elicit evidence of core ideas.  Given that these tests were designed to assess conceptual learning, this is what we would expect to find. This also suggests that the 3D-LAP is capable of identifying the kinds of questions that assess important concepts in physics. Crosscutting concepts are assessed significantly less frequently than the core ideas and scientific practices are almost never assessed by these concept inventories. This suggests that concept inventories are assessing students knowledge about physics concepts, but not necessarily their ability to do physics with those concepts. 

\paragraph{FCI} Our coding of the FCI demonstrates that few items have the potential to engage the student with more than one dimension. In fact, no items of the FCI were coded as three dimensional (Fig.~\ref{figure:nDLpoints}). Most of the points that can be awarded to students on the FCI are for 1-dimensional questions (63\%). There are a small fraction of points awarded for 2-dimensional questions (17\%) with the rest of the points awarded for answer questions with no dimensions (20\%). A close look at Fig.~\ref{figure:3DLpoints} shows why this is the case, 73\% of the points can be awarded for questions focused on Core Ideas. Only a minority of the points are awarded for answering questions that can elicit a Crosscutting Concept (17\%) or Scientific Practice (7\%), so there is very little chance of overlap between the dimensions.

\paragraph{FMCE} The FMCE provides more evidence of 3DL than the FCI (Fig.~\ref{figure:nDLpoints}). A small fraction of points on the FMCE (9\%) are awarded for answering 3-dimensional questions and the majority of points awarded on the FMCE are available for answering 2-dimensional questions (51\%). The FMCE has few points awarded for questions with no dimensions (6\%), but a fair percentage for 1-dimensional questions (34\%). Fig.~\ref{figure:3DLpoints} illustrates that the larger percentage of points for 2 and 3-dimensional questions stem from the greater number of points allotted to assessing Crosscutting Concepts (60\%) and Science Practices (28\%) -- leading to greater overlap with the Core Ideas (74\%).

\paragraph{CSEM} For the CSEM (Fig.~\ref{figure:nDLpoints}), we again find the greatest number of points available is allotted to 1-dimensional questions (69\%). Questions with no dimensions (6\%) and 3-dimensional questions (3\%) comprise a minority of the test. Nearly one-quarter of points (22\%) are available for answering 2-dimensional questions. We find that the majority of points available (94\%) appear on questions that contain a Core Idea (Fig.~\ref{figure:3DLpoints}). This result coupled to the low percentage of questions containing a Crosscutting Concept (22\%) or Scientific Practice (6\%) explains the large number of 1-dimensional questions on the CSEM.

\paragraph{BEMA} The BEMA is quite similar to the CSEM and FCI (Fig.~\ref{figure:nDLpoints}) in that it has a large fraction of points allotted to 1-dimensional questions (67\%) with few points given for answers to zero dimensional (13\%), two-dimensional (13\%), and three-dimensional questions (7\%). This result is explained similarly to the CSEM by the observation that the majority of points available on the BEMA are for answering questions with a Core Idea (83\%) while the points available for answering questions aligning with a Crosscutting Concept (17\%) and Scientific Practice (13\%) are low (Fig.~\ref{figure:3DLpoints}).

\paragraph{Comparing common assessments} The FCI and FMCE are often used in introductory courses to test students' conceptual understanding of classical mechanics. We have found that these assessments differ in the degree to which they assess three-dimensional learning. In fact, a contingency table analysis of this result shows that the frequency of tasks aligning with 0, 1, 2, and 3 dimensions is notably different between the two exams ($\chi^2 = 42.2$, $p \ll 0.05$, $\nu = 3$). We interpret this as suggestive that the FMCE is a better, albeit incomplete, measure of three-dimensional learning in physics when compared to the FCI. We find a similar, but not quite significant, association for the CSEM and BEMA ($\chi^2 = 6.5$, $p = 0.08$, $\nu=3$). However, here it is less clear which may be the better measure of 3DL, as the CSEM has a higher percentage of points aligning with crosscutting concepts and core ideas, while the BEMA has a higher percentage aligning with scientific practices.

\paragraph{Presence of specific components of 3DL}
While the analysis above provides an indication of the presence or absence of the potential to elicit evidence of a student engaging with scientific practices, crosscutting concepts, and core ideas, identifying the specific components that appear in each concept inventory requires that we delve more deeply into the coded data. Here, we identify which scientific practices, crosscutting concepts, and core ideas appear on each concept inventory at least once. Through this analysis we aim to answer our second research question: For which of the Scientific Practices, Crosscutting Concepts, and Core Ideas, do these concept inventories provide some evidence of student learning?

Table~\ref{table:DimensionsList} lists which dimensions appear at least once on each of the concept inventories. Only three (of seven) scientific practices, three (of seven) crosscutting concepts, and three (of five) core ideas are potentially assessed by these four concept inventories.  Within the scientific practices, ``Using Mathematics and Computational Thinking'' came up in three of the concept inventories.  The crosscutting concept of Scale, Proportion, and Quantity appeared on all four, and the core idea of Interactions Can Cause Changes in Motion appears on all of them.

\begin{table*}[htbp]
\caption{The scientific practices, crosscutting concepts, and core ideas that are potentially being assessed by each of the four concept inventories.\label{table:DimensionsList}}
\begin{tabular}{p{1cm}|p{5.5cm}|p{4cm}|p{5.5cm}}
&Scientific Practices &Crosscutting Concepts &Core Ideas \\
\hline
FCI   & Analyzing and Interpreting Data & Scale, Proportion, and Quantity & Interactions Can Cause Changes in Motion\\
FMCE  & Using Mathematical and Computational Thinking\newline Constructing Explanations and Engaging in Argument from Evidence & Scale, Proportion, and Quantity\newline Stability and Change & Interactions Can Cause Changes in Motion\newline Energy is Conserved \\
CSEM  & Using Mathematical and Computational Thinking & Scale, Proportion, and Quantity & Interactions Can Cause Changes in Motion\newline Interactions are Mediated by Fields\newline Energy is Conserved \\
BEMA  & Using Mathematical and Computational Thinking\newline Constructing Explanations and Engaging in Argument from Evidence & Scale, Proportion, and Quantity\newline Structure and Function & Interactions Can Cause Changes in Motion\newline Interactions are Mediated by Fields \\
\end{tabular}
\end{table*}

\section{Conclusion \& Discussion}\label{conclusion}




We used the lens of three-dimensional learning to analyze four of the most common concept inventories used by the physics community to see how well they can assess both students' knowledge of physics concepts and students' abilities to use those concepts to do physics \cite{national_research_council_framework_2012, national_research_council_discipline-based_2012}.  Using the 3D-LAP, we found that almost all of the tasks on these assessments align with at least one of the three dimensions originally defined by the {\it Framework}, but very few align with all three ($<$10\% on each concept inventory). Further analysis suggests that the alignment with dimensions is biased towards traditional conceptual goals, with evidence of eliciting the core ideas being much more common ($>$70\% on each) than the scientific practices ($<$25\% on each). Evidence of the crosscutting concepts being elicited was also low, though the FMCE does have notably more tasks aligned with crosscutting concepts than the other three conceptual inventories.

Each concept inventory did align with each of the three dimensions on at least one task. Across the four concept inventories, three of them included at least one task that aligned with the scientific practice of Using Mathematics and Computational Thinking. All four contained at least one task that aligned with the crosscutting concept of Scale, Proportion, and Quantity and the core idea of Interactions Can Cause Changes in Motion.

While our analysis reveals a number of shortcomings with the most widely-used assessments for introductory physics, the work is not without shortcomings. Analyzing these concept inventories using the 3D-LAP means we are looking at whether or not the tasks align with each of the dimensions of 3DL and almost nothing else. We take for granted that it is important for students to be assessed on both the practices and concepts of physics. We do not analyze aspects such as how students interpret the questions, the context in which the assessments are given, or other ways to analyze questions that are known to influence how students respond to them such as bias and readability \cite{national_research_council_knowing_2001}.

Nevertheless, these results suggest that concept inventories are not productive for gathering evidence of student learning that aligns with three-dimensional learning, particularly with regard to scientific practices and crosscutting concepts.  Again, our goal here is not to disparage concept inventories; they were designed to measure students' conceptual understanding and not to align with three-dimensional learning. Our goal was to determine how productive these existing assessments are from the lens of assessing three-dimensional learning, which came along later. This study suggests that there is room for improvement when it comes to aligning standardized assessments in college level physics with modern learning goals such as engagement in scientific practices. Further, this study suggests that the ability of concept inventories to obtain evidence that students are meeting modern learning goals are tenuous at best.

As discussed in Sec.~\ref{standardized}, concept inventories have played a vital role in changing the way introductory physics courses are taught and the curricula used for those courses.  However, another perspective is that the changes to curriculum and instruction that have proliferated in physics education would not have succeeded if they did not improve students' scores on concept inventories. It is hard to imagine any of these reforms being successful if the students' gains on the relevant concept inventory were lower in the new environment than in a traditional environment. In the PER community, researchers have developed other methods to investigate student learning as part of their research (e.g. affective measures, interviews, etc.), which might temper this sentiment, but for traditional physics faculty who use these assessments, we may be driving them toward ``maximizing'' a kind of learning that does not necessarily align with our modern understanding of what we want students to learn \cite{national_research_council_framework_2012,national_research_council_discipline-based_2012}. It is important to improve standardized assessments in the near future because they can drive curricular and pedagogical change in physics and, thus, have a significant impact on student learning at a large scale.

In the future, we aim to develop standardized assessments that align more fully with three-dimensional learning. Such assessments should be more capable of assessing students' abilities to use the centrally important ideas of physics to model, investigate, analyze, predict, and explain real world phenomena.  Additionally, we intend that such assessments communicate to the larger physics community that our learning goals are shifting to include both concepts and practices.

\acknowledgments{The authors would like to thank Kansas State University's Department of Physics and the Association of American Universities' STEM Education Initiative for their support. Additionally, we would like to thank the DBER community at Michigan State University (especially those involved in the development of the 3D-LAP).}

\bibliography{papers}  	

\end{document}